\documentclass[twocolumn,aps,superscriptaddress,pre]{revtex4}

\usepackage{amsmath,amssymb,graphicx}
\usepackage{algorithmic}
\usepackage{enumerate}
\usepackage{times}
\usepackage{color}
\usepackage{soul}
\definecolor{yblue}{rgb}{0.06, 0.3, 0.57}
\usepackage[pdftex]{hyperref}
\hypersetup{colorlinks=true,linkcolor=blue,citecolor=blue,urlcolor=blue}

\begin{document}

\title{Pinning effects in a two-dimensional cluster glass}

\author{Wenlong Wang}
\email{wenlongcmp@scu.edu.cn}
\affiliation{Department of Physics, Royal Institute of Technology, Stockholm, SE-106 91, Sweden}
\affiliation{College of Physics, Sichuan University, Chengdu 610065, China}

\author{Rogelio D\'iaz-M\'endez}
\affiliation{Department of Physics, Royal Institute of Technology, Stockholm, SE-106 91, Sweden}

\author{Mats Wallin}
\affiliation{Department of Physics, Royal Institute of Technology, Stockholm, SE-106 91, Sweden}

\author{Jack Lidmar}
\affiliation{Department of Physics, Royal Institute of Technology, Stockholm, SE-106 91, Sweden}

\author{Egor Babaev}
\affiliation{Department of Physics, Royal Institute of Technology, Stockholm, SE-106 91, Sweden}

\begin{abstract}
We study numerically the glass formation and depinning transition of a system of two-dimensional cluster-forming monodisperse particles in presence of pinning disorder. The pairwise interaction potential is nonmonotonic, and is motivated by the intervortex forces in type-$1.5$ superconductors. Such systems can form cluster glasses due to the intervortex interactions following a thermal quench, without underlying disorder. We study the effects of vortex pinning in these systems. We find that a small density of pinning centers of moderate depth has limited effect on vortex glass formation, i.e., formation of vortex glasses is dominated by intervortex interactions. At higher densities pinning can significantly  affect glass formation. The cluster glass depinning, under a constant driving force, is found to be plastic, with features distinct from non-cluster-forming systems such as clusters merging and breaking. We find that in general vortices with cluster-forming interaction forces can exhibit stronger pinning effects than regular vortices.
\end{abstract}

\maketitle

\section{Introduction}
Since the past few decades there have been significant ongoing efforts directed to understanding the rich properties of vortex matter in type-2 superconductors \cite{blatter:94}. 
The glass formation has been intensively studied for vortex systems with pinning disorder \cite{VG1,VG2,blatter:94,VG4}. 
One major research theme is investigating the effects of pinning disorder on vortex glass formation, and the corresponding depinning transition and dynamics upon driving. This is important in superconductor applications, as dissipationless current-carrying states require arresting the dynamics of vortices.
In the much less studied multicomponent superconductors, 
the vortex interactions
can take complex forms as a result of multiple competing 
(attractive and repulsive) interaction length scales \cite{Babaev.Speight:05,Babaev.Carlstrom.ea:10,carlstrom2011type,Silaev.Babaev:11,babaev2017type}. 
This has been proposed not only in 
superconductors with several superconducting components but also in multilayer systems with different coherence ($\xi_i$) and penetration ($\lambda_i$) lengths \cite{carlstrom2011semi,varney2013hierarchical,Rogelio:Glass}. 
In the bulk of intrinsic type-$1.5$ superconductors, intervortex forces are long-range attractive and short-range repulsive \cite{Babaev.Speight:05,Babaev.Carlstrom.ea:10,Silaev.Babaev:11,babaev2017type}. Various multiband materials were proposed to fall into the type-1.5 class \cite{RB1,ray2014muon,PhysRevB.102.144523}.
The double-transition materials are good candidates for type-1.5 behaviour due to
the diverging coherence length at the lower transition  \cite{carlstrom2011length,garaud2018properties}.
For thin films, an additional repulsive interaction arises due to the magnetic stray fields.
Many vortex phases have been predicted for vortices in type-$1.5$ superconducting systems \cite{meng16,meng2014honeycomb,varney2013hierarchical,Rogelio:Glass}. Particularly, the cluster phases are in many ways similar to ones
emerging in a wide range of cluster-forming monodisperse particle systems such as colloidal suspensions and ultracold atoms \cite{MPstripe,cinti14,pc,RR:stripes,RR:phases,RR:depinning,ca:patterns,CCSH,Glaser:SS}.

In type-2 non-cluster-forming vortex systems, glass phases are associated with pinning, while
in contrast, with cluster-forming interactions the situation is different and 
quench dynamics can produce a glass also without pinning \cite{Rogelio:Glass}. The effect of pinning is much less understood for type-1.5 superconductors. The especially interesting  question is the effect of pinning on vortex cluster glass formation. In addition, the nonequlibrium depinning transition of monodisperse particles has been intensively studied for particle solids; see \cite{RR:depinning} for a recent review. 
However, the depinning of pattern forming systems, in particular the cluster phases \cite{RR:hysteresis} which are experimentally relevant, has not been extensively studied. Indeed, even the nature of the depinning and the main features of the flowing phase have not been explored in detail.

The main purpose of this work is to present a systematic study of cluster glass formation following a thermal quench in presence of pinning disorder, and the corresponding depinning dynamics when a constant external force is applied. We are interested in, e.g., how pinning disorder density and depth affect glass formation, and in fact whether cluster glass itself remains well defined, and for depinning we are interested in the nature of the depinning transition, i.e., elastic depinning or plastic depinning, and the main features of the flowing phase. Here, plastic flow refers to that the flowing clusters change their neighbours, while in elastic flow, they do not.

For traps of a moderate depth, we find that cluster glass remains reasonably well defined, and persists even to the regime of dense pinning.
The depinning transition is found to be plastic, and the flowing phase has a number of interesting features compared with the non-cluster-forming counterpart.
In addition to the cluster plastic flow, the clusters can undergo structural transformations. 
Most notably, the clusters can merge or break up due to thermal effects, 
and, more importantly, the interplay between the pinning force and the driving force. 
The effective polydispersity assists the plastic flow also in that a smaller cluster tends to have a better mobility, and consequently diffuses and escapes from a trap faster. Finally, we have compared the depinning of a cluster glass and a particle glass, finding that the cluster glass exhibits stronger pinning effects. This suggests that cluster vortex glasses might be suitable for certain technological applications where a high critical current is desired.

The paper is organized as follows: the model, observables, and numerical methods are described in Sec.~\ref{mm}. 
Numerical results are given in Sec.~\ref{results}. 
Finally, a summary and discussion of the main findings is presented in Sec.~\ref{cc}.

\section{Model, observables, and methods}
\label{mm}

\subsection{Model}

We study a system of two-dimensional monodisperse particles interacting via an effective pairwise potential, 
and with pinning centers, and an external driving force modelled by the following dimensionless Hamiltonian:
\begin{eqnarray}
H = \sum_{i<j} U_{ij} - \sum_{i\alpha} U_p \exp[-r_{i\alpha}^2/(2\sigma^2)] - F_D\sum_i x_i.
\end{eqnarray}
The indices $i, j$ run from 1 to the number of particles $N$, and 
$\alpha$ runs from 1 to the number of pinning centers $M$. 
The three terms are for particle-particle interactions, randomly located and quenched Gaussian 
pinning centers, 
and 
a constant driving force along the $x$-direction, respectively.
The particles are placed on an $L \times L$ square with periodic boundary conditions. The density of particles is $n=N/L^2$ and the density of impurities is $m=M/L^2=nM/N$. 
$U_p>0$ and $\sigma$ denote the depth and size of the pinning centers, respectively. 
$F_D$ denotes the strength of the driving force. We 
used $\sigma=0.7$ 
unless specified otherwise, which is approximately half of the lattice constant of the corresponding cluster crystal 
in the absence of disorder and driving.
A point-particle representation is applicable for superconducting vortices where
the interparticle interaction forces are chosen to be consistent with those
obtained in superconductivity models.

\begin{figure}[htb]
\begin{center}
\includegraphics[width=\columnwidth]{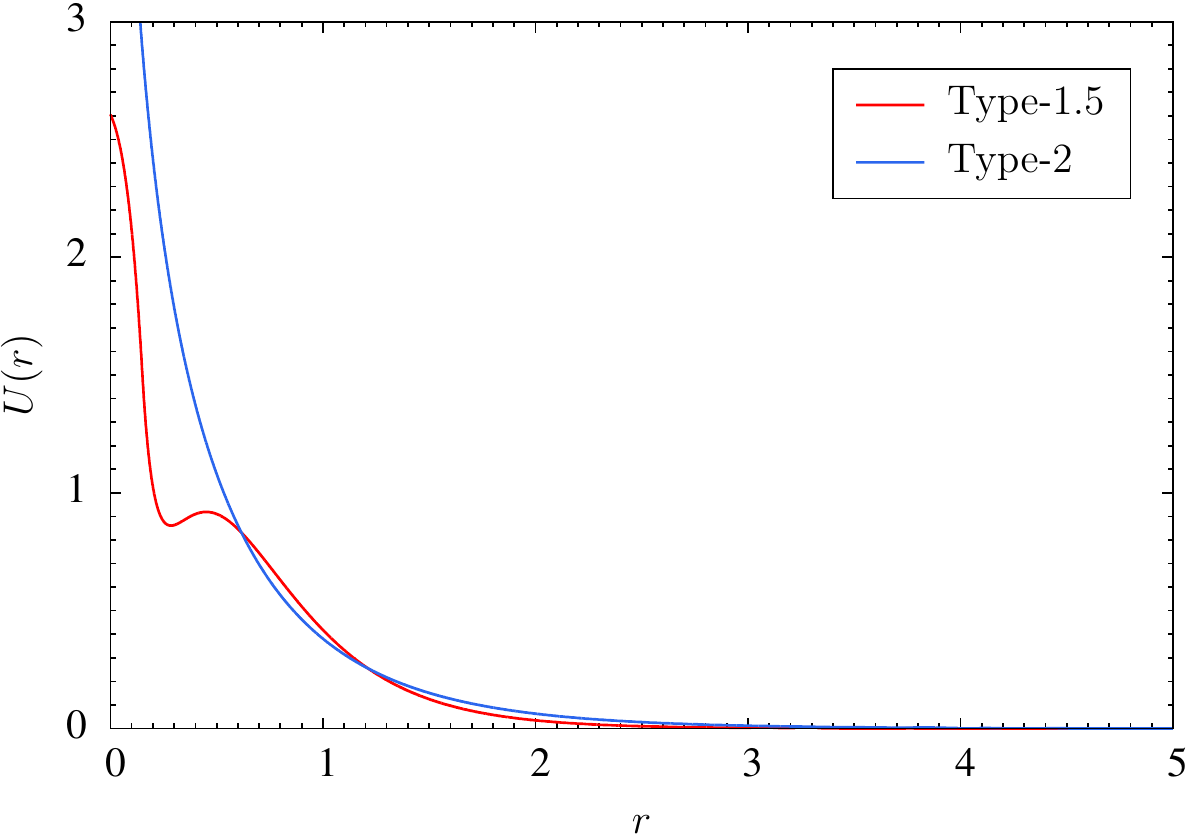}
\caption{
Pairwise particle-particle, i.e., vortex-vortex interaction potentials.
The red curve is a cluster-forming potential [Eq.~(\ref{potential})] while the blue curve forms single-particle crystals. 
They model vortex interactions
in type-1.5 and type-2 superconductors, respectively.
}
\label{UR}
\end{center}
\end{figure}

The potential, relevant to vortex interactions in 
type-1.5 superconductors, is defined as: 
\begin{eqnarray}
        U(r) =
        \begin{cases}
        U_0 - \alpha_0(r+0.15)^4, & \text{if } r\leq0.15, \\
        \sum_{l=1}^{3} C_lK_0(\alpha_l r), & \text{if } r>0.15,
        \end{cases}  
    \label{potential}  
\end{eqnarray}
where $K_0$ is the modified Bessel function of the second kind, 
and the parameters are 
$U_0=2.6796, \alpha_0=144.3760, C_1=7.2900, \alpha_1=2.3810, C_2=37.2100, \alpha_2=7.1174, C_3=-36.1911, \alpha_3=5.1020$. 
The potential is plotted as the red curve in Fig.~\ref{UR}, it is similar to the one used in \cite{Rogelio:Glass} in medium and long ranges, but it features an attractive intermediate part and the unphysical divergence from long-range asymptotic analysis at the origin is regularized with a fourth-order polynomial to match better with the full numerically calculated potential \cite{carlstrom2011semi}.

For comparison, we also use the standard repulsive type-2 vortex 
interaction potential from a single modified Bessel function $U(r)=CK_0(\alpha r)$ with parameters $C=1.7511$ and $\alpha=1.4867$, 
shown as the blue curve in Fig.~\ref{UR}.
This potential generates single particle crystals with no clustering behavior. 

The physical origin of the potential (\ref{potential}) is the following.
In the type-1.5 regime a superconductor has several coherence lengths,
originating from the individual superconducting components. Some of the coherence lengths are
larger and some are smaller than the magnetic field penetration length.
The resulting intervortex forces asymptotically have the form
of a sum of Bessel functions 
which give contributions to attractive interactions
at the coherence length scales 
and repulsive interaction at the magnetic field 
penetration length \cite{Babaev.Speight:05,Babaev.Carlstrom.ea:10,babaev2017type,carlstrom2011type,garaud2018properties}.
In the standard type-1.5 regime the intervortex interaction is short-range 
repulsive and long-range attractive.
However, the situations can also occur where the interaction has a local minimum at an intermediate length scale.
This is expected to be realized in 
thin films of type-1.5 superconductors where
at the long range the interaction should be repulsive due to stray fields \cite{Rogelio:Glass},  
and in layered systems where there are different
magnetic field penetration lengths in different layers \cite{varney2013hierarchical,meng16,Rogelio:Glass}. Multiple repulsive length scales can also arise as an intrinsic feature in certain anisotropic models \cite{winyard2019hierarchies}.

Since we are interested 
in not very dense ensembles the only retained attractive Bessel function
corresponds to the interactions of larger cores.
We consider a bilayer system with different penetration lengths that 
is approximated by two repulsive Bessel functions. Note that the 
potential with such an intermediate minimum also arises 
in a thin film of type-1.5 superconductors where instead of the Bessel function with the largest length one includes a power-law repulsion \cite{Rogelio:Glass}.
We emphasize that this does not affect the existence of cluster crystals, which appear at moderately low densities.
Such a regime is the main focus of this work.

We briefly compare our potential to other commonly used cluster-forming potentials.
In addition to \cite{Rogelio:Glass}, our potential is also similar to ultrasoft 
potentials such as $U(r)=1/(1+r^6)$ \cite{Wang:VC} in the intermediate and long ranges, but differs near the origin. 
Therefore the resulting phase diagrams differ at high densities, but all three potentials lead to cluster crystal phases at low densities.
Our potential has a characteristic
medium attractive range from the attractive Bessel term.
This feature is similar to the potential studied in \cite{RR:phases}, which is a sum of a repulsive Coulomb term and an attractive exponential term. The two potentials again differ near the origin and therefore the high-density regimes will be different.

\subsection{Observables and methods}

Both equilibrium and dynamical properties of the system were studied for a large range of particle as well as trap densities
using Monte Carlo (MC) and molecular dynamics (MD) simulations.
The simulations have three different parts. 
First, equilibrium sampling was used to map the phase diagram of the clean system with no pinning and no driving, i.e., $U_p = F_D =0$. 
The transition temperatures as well as the ordered low-temperature phases
were obtained using parallel tempering MC \cite{ptmc1,ptmc2,Hukushima:PT}.
The calculations were complemented by simulated annealing MC \cite{SA} as a consistency check.
While simulated annealing cannot faithfully maintain thermal equilibrium at low temperatures, 
it can nevertheless capture the essential features of the phases, e.g., 
produce typical low energy states \cite{Wang:VC,ca:patterns}. 
We used the Metropolis algorithm with a sequential update order of the particles.

For the glass formation simulations with pinning but no driving,  
i.e., $U_p \neq 0$ and $F_D=0$, we have used single temperature MC simulations, and 
a random update order of the particles
starting from fully disordered configurations.

For the depinning transition, both pinning and driving are included, i.e., $U_p \neq 0$ and $F_D \neq 0$. 
Since it is unclear if MC can correctly capture transport dynamics, we 
instead used MD for this study.
Details of the MD simulation of the overdamped Langevin dynamics 
are discussed in Sec.~\ref{dt}.

The MC simulation methods used here are essentially the same as in \cite{Wang:VC}. 
Each MC sweep for a replica is $N$ attempted updates of the particles in either sequential or random order. 
The MC trial moves are attempts to shift the particle position randomly within a square box of length $2\sqrt{1/n}$ centered on the particle. We only use single-particle update moves.
For glass formation dynamics this corresponds to the overdamped regime.

The
main observables are the cluster orientational order parameter $\phi_6$, the 
heat capacity $C_V=\beta^2 \rm{var}(E)$ where $E$ is the total energy of the system, and the pair correlation function $g(r)$.
The parameter $\phi_6$ quantifies the ordering of the triangular cluster-crystal phase. 
The calculation of $\phi_6$ for a given configuration is based on the cluster positions 
and we now discuss in detail how to compute it.

We first use a hierarchical clustering technique \cite{DataMining,Rogelio:Cluster} that group particles into clusters deterministically. 
Each particle starts as a single cluster, and the two closest clusters are joined together if their center of mass distance is smaller than a chosen cutoff.
We used a cutoff of $0.7 \approx a/2$, where $a$ is the lattice constant of the corresponding cluster crystal in the absence of pinning disorder; see Fig.~\ref{PD}. 
The results present in this work are not sensitive to the precise value of the chosen cutoff, e.g., around $0.7\pm 0.1$, as we work in parameter regimes where clusters are reasonably well defined. 
When two clusters 
are merged, the new center of mass is updated. 
The process is repeated until no further grouping is accepted. 
Hence the process takes a particle configuration $\{\vec{r}_i, i=1,2,...,N\}$ 
and outputs the centers of mass $\{\vec{R}_i, i=1,2,...,C\}$ of the $C$ clusters. 
The clustering process is followed by a 
Voronoi decomposition to identify neighbouring clusters.

The orientational order parameter $\phi_6$ is finally defined as:
\begin{eqnarray}
\phi_6 = \left| \dfrac{1}{C} \sum_{j=1}^{C} \left( \dfrac{1}{N_j} \sum_{\ell=1}^{N_j} e^{i6\theta_{j\ell}} \right) \right|,
\end{eqnarray}
where $\theta_{j\ell}$ is the angle 
between the vector $\vec{R}_{\ell} - \vec{R}_j$ and an arbitrary direction, often the $\hat{x}$-axis, 
and $N_j$ is the number of neighbours of the $j_{th}$ cluster. 
Note that only neighbouring pairs are summed over. 
For a perfect triangular lattice, $\phi_6=1$, and with no long-range order and random positions, $\phi_6=0$.
Hence $\phi_6$ is an order parameter. 

The pair correlation function is defined as:
\begin{eqnarray}
g(r) = \frac{1}{N} \dfrac{\delta n(r)}{2\pi r \delta r},
\end{eqnarray}
where $\delta n(r)$ is the number of particles in the shell $2\pi r \delta r$, with reference to an arbitrary particle. 
Note that the function is normalized as $\int_0^{\infty} g(r) 2\pi r dr =1$. 
The lattice constant of a cluster configuration can be extracted from the peaks of $g(r)$.

\section{Numerical results}
\label{results}

\subsection{Cluster glass formation}

\begin{figure}[htb]
\begin{center}
\includegraphics[width=\columnwidth]{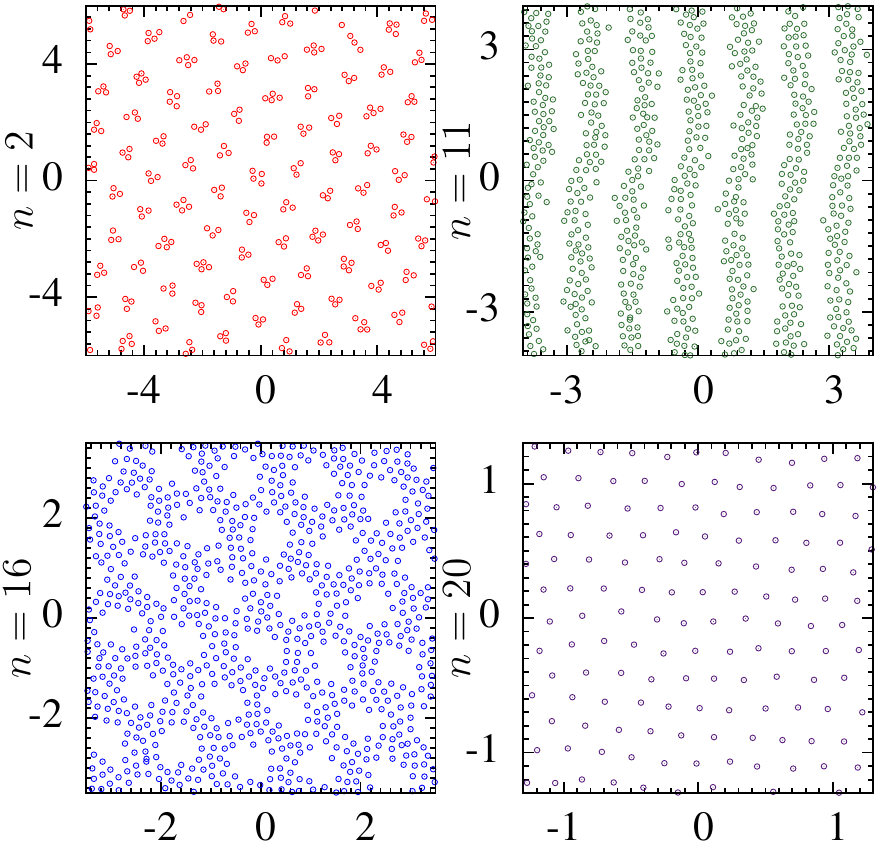}
\includegraphics[width=\columnwidth]{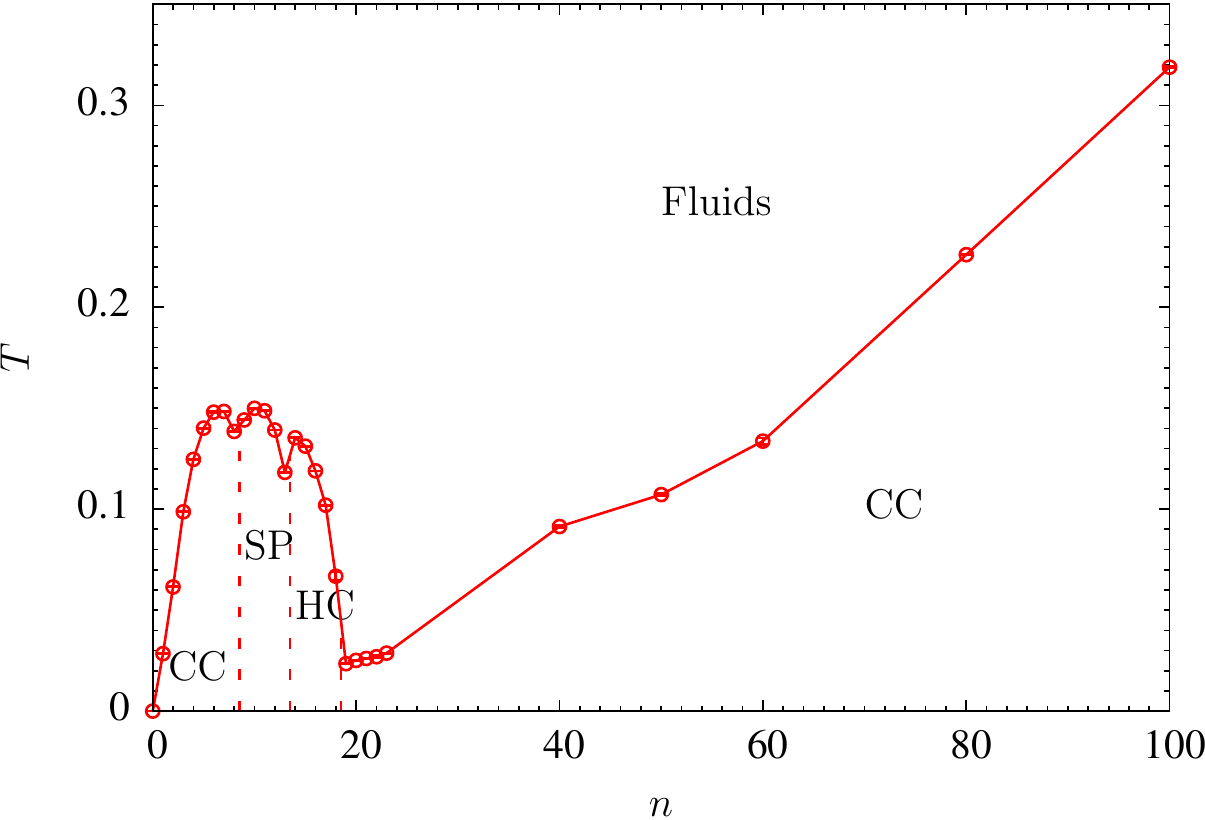}
\caption{
Top panels: Typical equilibrium configurations of the ordered phases of cluster crystals (CC), stripes (SP), hole crystals (HC) and simple crystals in order of increasing density.
Bottom panel: The schematic phase diagram of the potential [Eq.~(\ref{potential})] in absence of pinning disorder estimated from the major peak of the specific heat. The two cluster crystal phases have clustering onset densities at about $n_c=0.65$ and $30.5$, respectively. 
}
\label{PD}
\end{center}
\end{figure}

In order to study cluster glass formation, it is important to study the underlying pure system without pinning disorder for input, particularly the relevant particle density for clustering and the corresponding transition temperature. The phase diagram of the pure system along with all the ordered phases are depicted in Fig.~\ref{PD}. Here, the transition temperature is estimated from the prominent $C_V$ peak using $N=1000$, which is sufficiently large for a reasonably accurate estimation. The crystalline phases are identified by examining typical low-temperature equilibrium configurations, and they are also verified from ``ground states'' found by simulated annealing. It should be noted that our phase diagram here is only a sketch, and the more subtle features such as the number and the order of these phase transitions are not considered. For the potential in Eq.~(\ref{potential}), we find ordered phases of cluster crystals, stripes, and hole crystals as the particle density is increased up to about $n=18$. This series of ordered phases appears to be rather common in cluster-forming potentials with a repulsive core \cite{RR:phases,Rogelio:Glass}. Note that our potential restores the cluster crystal phase at even higher densities after the hole crystal phase.  This behaviour is different from the potentials with divergent cores near $r=0$; cf. \cite{RR:phases,Rogelio:Glass}. It should be noted that this second cluster phase at high density differs from the first cluster phase at low density with, e.g., a very distinct lattice constant. While pattern formations and the nature of these phase transitions are interesting in their own right, we here focus on the glass formation of the physically relevant low-density cluster phase.

We now present more details of the first triangular cluster phase. The onset density of clustering is at about $n_c=0.65$, below which there is only one particle per unit cell. When clustering occurs, the lattice constant settles to $a \approx 1.34$, which is estimated from the pair correlation function, and is independent of the density. Interestingly, both the onset density and the lattice constant are very similar to that of the ultrasoft potential $U(r)=1/(1+r^6)$ \cite{Wang:VC}. 
This is presumably because the two potentials have in some sense similar features in the medium and long ranges. In the following, we work with a typical cluster phase at density $n=2$ where each cluster has approximately $3$ or $4$ particles. The transition temperature at this density is $\beta_c \approx 16.3$. We now turn to the cluster glass formation of this system.

In the absence of pinning disorder, clustering particles form cluster glasses when quenched from a high temperature, e.g., $T=\infty$ to a sufficiently low temperature in a single step \cite{Rogelio:Glass}. The system relaxes following the quench, and it was found that the system can reasonably restore the equilibrium order parameter except when the final quenching temperature $T_f$ is too low. The glass transition temperature is then loosely defined when the resulting order parameter $\phi_6$ departs from that of the equilibrium cluster crystal phase. It was argued in \cite{Rogelio:Glass} that cluster-forming particles are better glass formers compared with non-cluster-forming particles because of the effective polydispersity of the cluster sizes. On the other hand, it is well known that pinning by itself can lead to glass formation in conventional vortex matter \cite{blatter:94}. We next investigate the situation where both mechanisms are present.

For the simulation of thermal quench dynamics in presence of pinning disorder we consider different density and strength of pinning centers, and compare results with the case of no pinning. For each set of parameters we run quench dynamics to various final temperatures $T_f$ from 
random configurations corresponding to $\beta=0$, using single-temperature Monte Carlo dynamics.
For all the temperatures that we studied $10^5$ sweeps were enough for a good relaxation, and then
the order parameter $\phi_6$ of the final configuration was measured. 
Only one measurement per temperature was made from one disorder realization to prevent correlations, and statistics were collected using averages over disorder realizations. Here, $100$ realizations are simulated for each pinning density and strength.

\begin{figure}[htb]
\begin{center}
\includegraphics[width=\columnwidth]{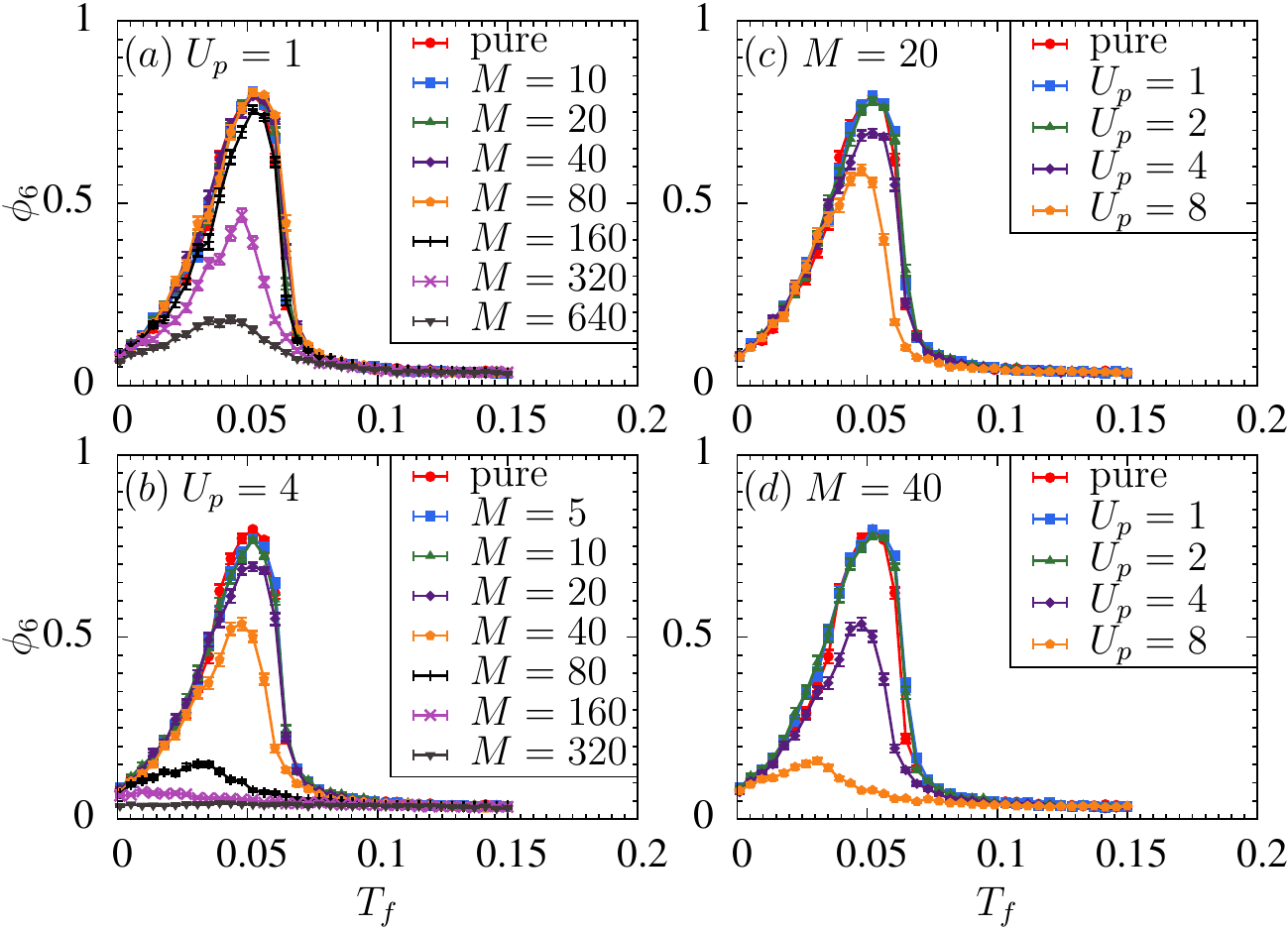}
\caption{
Effects of density and pinning strength on glass formation after a thermal 
quench as a function of the final temperature $T_f$ at density $n=2$ with $N=1000$ particles. 
The topmost red curve in each panel is the pure case without disorder which provides an upper bound. 
Left panels: Changing the number of pins
$M$ at fixed trapping strengths (a) $U_p=1$ and (b) $U_p=4$. 
Right panels: Changing the pinning strength 
at fixed number of pins for
(c) $M=20$ and (d) $M=40$.
}
\label{Dynamics}
\end{center}
\end{figure}

Results for dynamical glass formation are summarized in Fig.~\ref{Dynamics}. 
In all panels, the red (topmost) curve corresponds to thermal quench of the clean system. 
This result is similar to that of the molecular dynamics for other cluster-forming potentials \cite{Rogelio:Glass}. 
There are two relevant temperature scales, one is the transition temperature of the pure system where 
the system forms global order as temperature is lowered below $T_C$. This matches well with the transition 
temperature we find from the 
heat capacity. 
The other temperature is the crystal-to-glass transition temperature $T_G$ \cite{Rogelio:Glass}, which can be estimated as where $\phi_6$ bends downwards and starts to drop, here, $T_G \approx 0.05$. 
This is a characteristic temperature below which dynamics becomes glassy and the system readily runs out of equilibrium. 

It is natural that the clean curve provides an upper bound for restoring the global order, since 
adding random pinning 
and increasing the pining strength generally 
enhance glass formation. 
However, it is interesting to note that the effect of pinning is far from linear. 
Pinning actually has little effect on the order parameter if the disorder is weak or dilute. For example, at $U_p=1$ there is no prominent change of the order parameter up to about $M=160$. Further increasing the number of pins or the pinning strength eventually makes the order depart prominently from the clean limit. In the strong pinning limit, the global order is essentially entirely eliminated. 
Another interesting observation that can be drawn from Fig.~\ref{Dynamics} is that addition of pinning gives the strongest suppression of the $\phi_6$ order
parameter at the intermediate temperatures, while at least for relatively dilute pinning the curves merge with the clean curve at lower
temperatures. This suggest that at least for dilute pinning the system falls into a state similar to
that of an interaction-dominated effective polydispersity class pinned by rare pinning centers.

\begin{figure}[htb]
\begin{center}
\includegraphics[width=\columnwidth]{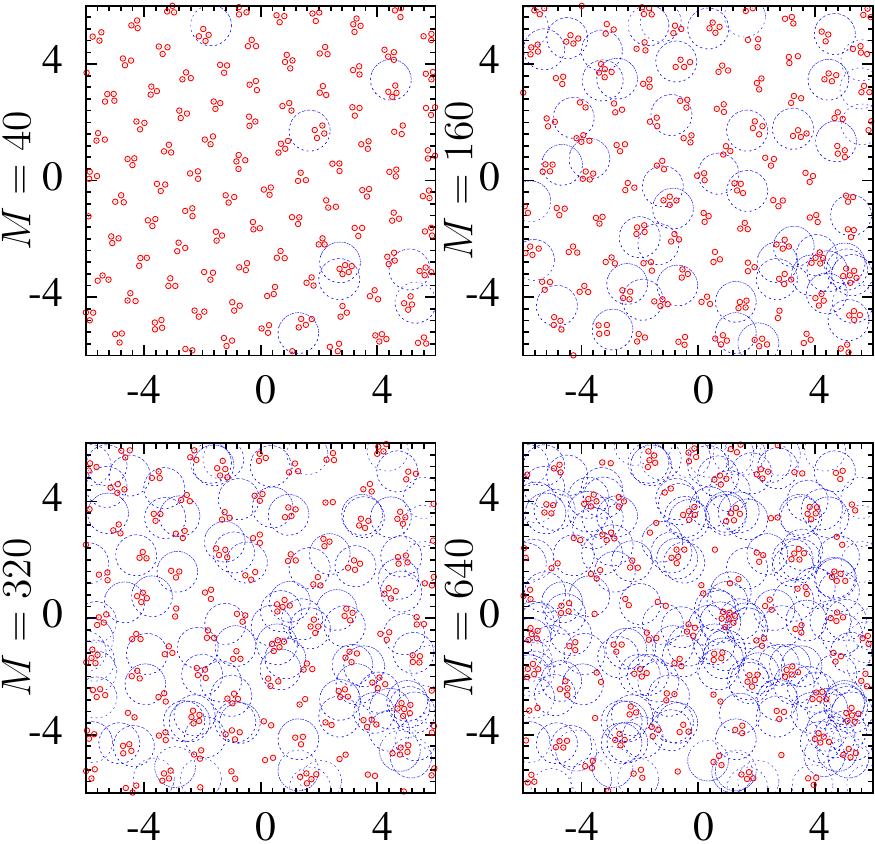}
\includegraphics[width=\columnwidth]{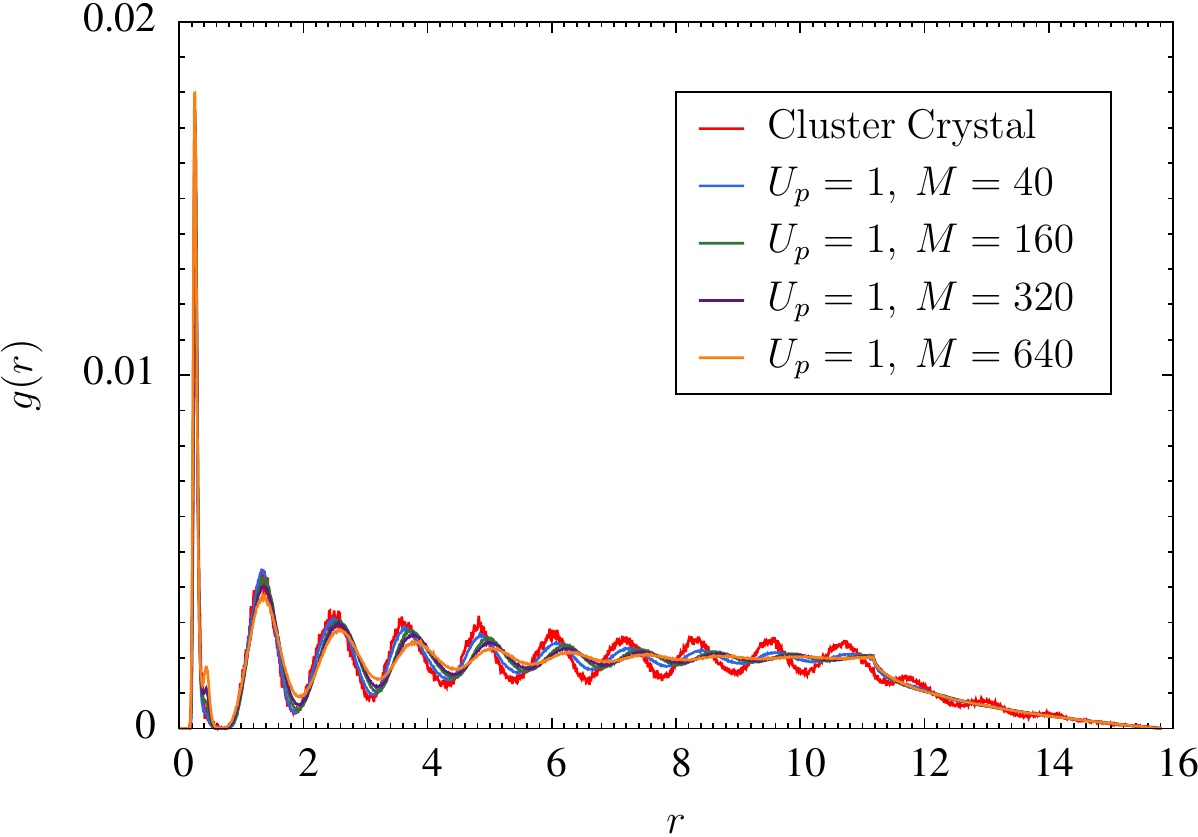}
\caption{
Top panels: Typical cluster glass configurations corresponding to those of Fig.~\ref{Dynamics} at approximately $T=0.03$. The blue dashed circles are attractive Gaussian centers. Note that a fraction of some large clusters is formed in or near the (overlapping) traps. These configurations suggest that clusters are reasonably well defined in our parameter regimes.
Bottom panel: The clustering feature is further confirmed in the pair correlation function $g(r)$. Particularly, the second largest peak is the inter-cluster peak, or the lattice constant peak of the corresponding cluster crystal. The decay tail ($r \gtrsim 11)$ is a finite-size effect. 
}
\label{XY4}
\end{center}
\end{figure}

It is important to confirm that the quenched glasses are genuine cluster glasses in presence of substrate disorder. To this end, we have checked the final configurations. Several typical cases are illustrated in the top panels of Fig.~\ref{XY4}. The blue circles are Gaussian traps, it is clear from observation that clusters are still reasonably well defined, despite that the attractive traps distort the cluster order as well as the distribution of the cluster sizes. For example, it is more likely to find larger clusters in or near the (overlapping) traps. Interestingly, these clusters do not necessarily all sit at the centers of the traps.  In the dense case $M=640$, there are a number of large clusters of sizes $5, 6$ and even $7$, leaving at the same time some small clusters in the trap interstitial voids.

Cluster glass manifests itself when the corresponding cluster crystal lattice constant peak is relevant in the pair correlation function. This is shown in the bottom panel of Fig.~\ref{XY4} for various $M$ values. It is seen that the second largest peak, which is the cluster crystal lattice constant peak, is present for all the cases studied. The one near $r=0$ is from the neighbouring particles within clusters, note that a small shoulder peak is developed for the dense cases $M=320$ and $640$, as some clusters are larger compared with the typical size of the free case. By contrast, the lattice constant of the particle triangular crystal $3^{-\frac{1}{4}}=0.7598$ at the same density is not (at least yet) relevant. 

It is important to note that we should not take the weak and strong traps here too literally, as in our cases cluster glasses are \textit{all} reasonably well defined. It is clearly possible to further increase the trap depth, and new phases, e.g., particle glass may appear. It is expected that fully particle glasses should form in certain trap settings, depending on the trap size, depth, and density. Our main conclusion of this section is that interaction-driven cluster glasses are stable with respect to inclusion of quenched disorder.

\subsection{Cluster glass depinning dynamics}
\label{dt}

In this section, we use molecular dynamics to study the 
depinning dynamics of the cluster glasses we obtained. We use the overdamped Langevin dynamics following the stochastic ordinary differential equations:
\begin{eqnarray}
\gamma\dot{\vec{r}}_i &=& -\nabla U_i + \sqrt{2T\gamma} \vec{\eta}_i(t) + F_D \hat{x},
\end{eqnarray}
where 
without loss of generality we set the friction coefficient $\gamma=1$. 
The Gaussian stochastic force $\eta$ has zero mean and are uncorrelated in time, $\langle \eta_{i\alpha}(t) \eta_{i\gamma}(t') \rangle = \delta(t-t') \delta_{\alpha\gamma}$, where $\alpha, \gamma = x, y$ denote components in different spatial directions. $U_i$ is the potential energy of particle $i$ interacting with other particles and the traps, and $F_D$ is applied along the $\hat{x}$-axis. 
For the thermal noise in each direction, it is straightforward to show from the correlation function that $\int_t^{t+dt} \sqrt{2T} \eta_{i\alpha}(t') dt' = \sqrt{2Tdt}\zeta$, where $\zeta$ follows the standard Gaussian distribution $n(0,1)$. 
In the large driving limit, we expect a drift velocity $v = \langle v_x \rangle = F_D$ as in the case of single particles \cite{RR:depinning}.
The force on a particle for the symmetric potential can be evaluated exactly using the relations $F(r) = -\frac{dV}{dr} \hat{r}$ and $d K_0(r)/dr = - K_1(r)$ of the modified Bessel functions of the second kind. The $K_1$ function is similar in shape to $K_0$ but it is further shifted away from $r=0$. The integration is done using the fourth-order Runge-Kutta method. We study the cluster glass depinning for a typical case of $U_p=1$ at $T=T_C/2$, where various features of the cluster flow can be studied. In certain cases, we have also performed 
simulations at $T=0$ to suppress thermal fluctuations for a more clear visualization of the dynamics.

\begin{figure}[htb]
\begin{center}
\includegraphics[width=\columnwidth]{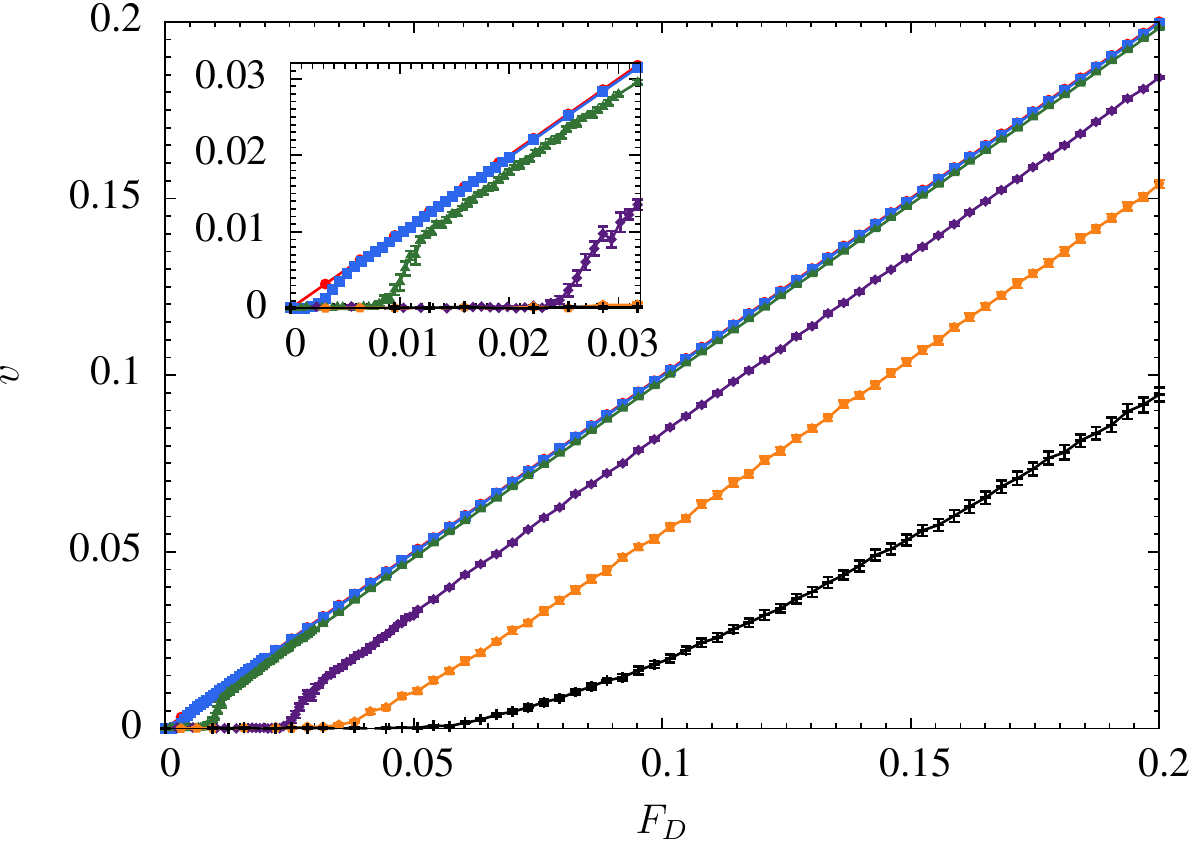}
\caption{
Drift velocity of the cluster glass subject to an external driving force at $T=T_C/2$ for, from left to right, the pure system, $M=10, 40, 160, 320$, and $640$, respectively.
The inset panel is a zoom of the main panel near the origin. 
For the pure case, there is a linear force-velocity relation. 
For dense pinning there is a prominent depinning transition and the drift velocity converges slowly to the free case. 
For 
dilute pinning the critical force is much smaller and the drift velocity converges rapidly to the pure case. The concave down of the depinning curves near the transition suggests plastic depinnings, which are subsequently confirmed in the details of the flowing dynamics; see the text for details. 
}
\label{Drive}
\end{center}
\end{figure}

In the pure 
case with no pinning, the clusters slide together along the direction of the driving force, and the drift velocity is given by the linear relation $v=F_D$ for any $F_D$; see the (topmost) red line of Fig.~\ref{Drive}. At finite temperatures, the particles will fluctuate while sliding, but the underlying cluster glass order is unchanged. When the temperature is quenched to $T=0$, the internal dynamics of the clusters will be rapidly frozen and the cluster glass essentially slides rigidly afterwards. In either case, the cluster glass remains intact.
These observations are expected since 
without pinning the particles motion is given by a
superposition of the internal dynamics when driving is absent and a constant translational motion, i.e., in the absence of pinning the state can be simply identified as a {\it sliding cluster glass}. 
A typical movie of such dynamics for $F_D=0.2$ at both finite and zero temperatures is presented in \cite{comment:mvmd1}. 

In presence of pinning, a dynamical depinning transition should occur as the driving force is increased. We have conducted a large-scale MD simulation to measure the disorder-averaged drift velocity at various driving forces. For each run, we integrate up to a sufficiently long time $t=1000$, and the drift velocity is measured using the latter half of the simulation, i.e., the initial transit dynamics is ignored and we measure the asymptotic steady state drift velocity. The results are shown in Fig.~\ref{Drive}. For moderate and dense pinnings, there is a prominent pinned phase. The critical depinning force is naturally very small but nonzero when the pinning is dilute. Detailed examination near $F_D=0$ reveals that there are pinned phases as well for dilute pinnings. 
For a depinning transition, a critical exponent $\beta$ can be defined as $v \sim (F-F_C)^\beta$ when $F \geq F_C$ in the vicinity of the depinning critical force $F_C$ \cite{Fisher:beta}. This is analogous to the order parameter scaling at a second order phase transition. As the curves are concave down near the transition, it suggests that $\beta>1$, i.e., the depinnings are likely plastic \cite{RR:depinning}. In plastic flows, particles or 
clusters change their neighbours during the flow, contrary to the elastic flow where neighbouring particles 
and clusters are preserved. A close examination 
of the dynamics shows that the cluster flows are indeed plastic, with a number of features distinct from the non-cluster-forming particle systems, e.g., clusters undergo merging and breaking dynamics.

\begin{figure}[htb]
\begin{center}
\includegraphics[width=\columnwidth]{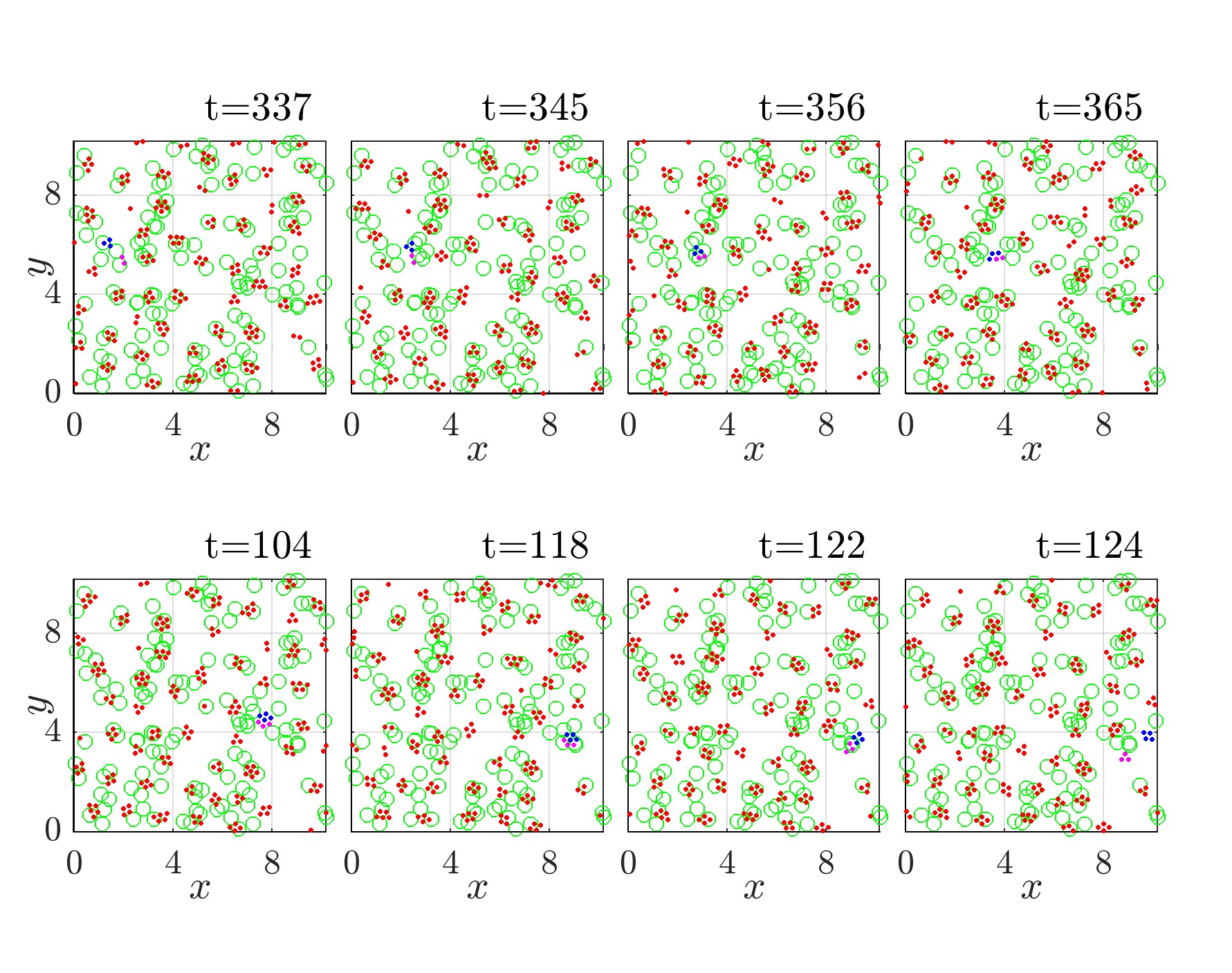}
\caption{
Typical clusters merging (top panels) and splitting (bottom panels) dynamics induced by the trap centers ($T=0$), the pertinent clusters are in blue and pink while other clusters are in red, and the green circles are pinning centers. In this dense pinning regime, such processes are frequent, rendering the flow highly plastic.
}
\label{pf}
\end{center}
\end{figure}

We discuss the dense pinning first, where the flowing phase is clearly plastic. The flowing dynamics is quite complex, and here we focus on the main features and typical dynamical processes. Considering the complexity of the flow, a movie showing both the pinned and flowing phases are shown in \cite{comment:mvmd2}, 
where the driving forces are $F_D=0.0508$ and $0.2$, respectively, for a disorder realization of $M=640$. There are a number of interesting observations for the flowing phase. First, cluster chaotic flow occurs instead of the particle chaotic flow in non-cluster-forming ensembles. Here, the clusters have a strong tendency to flow through or near the traps, forming rivers and changing neighbours. Interestingly, the pinning centers are like stepping stones for the clusters. In addition, the clusters are no longer as robust as in the free case. There are frequent clusters merging and breaking dynamics. For example, unusually large clusters may form as a result of merging and there are also a number of single-particle clusters as a result of 
particle emission from clusters. As a result, the clusters are much more heterogeneous than the pure case due to the interplay of pinning and driving. It is important to emphasize that this increased heterogeneity is \textit{not} a result of thermal noise as we are working in the low-temperature glass phase, despite that it also contributes. This is clearly demonstrated as the above processes also occur at $T=0$ as shown in the movie in \cite{comment:mvmd3}. A typical cluster merging process and also a cluster splitting process are depicted in Fig.~\ref{pf}. It is interesting that trap centers play a ``catalytic'' role in these dynamics. Indeed, these dynamics are much suppressed in the dilute pinning regime.

In the opposite dilute pinning regime, the flow is much more ordered. After close inspections of $M=10$, we find that the flow is still plastic, but it is very weak. 
Upon depinning the glass mostly slides over the pinning centers. The plastic flow and cluster merging and breaking are much suppressed. Nevertheless, the effective polydispersity of the clusters can still undergo trap-induced neighbour-changing dynamics because of their different mobilities. Similar to a cluster liquid \cite{Wang:VC}, small clusters tend to have better mobilities, and moreover larger clusters are better pinned by impurity traps.

\begin{figure}[htb]
\begin{center}
\includegraphics[width=\columnwidth]{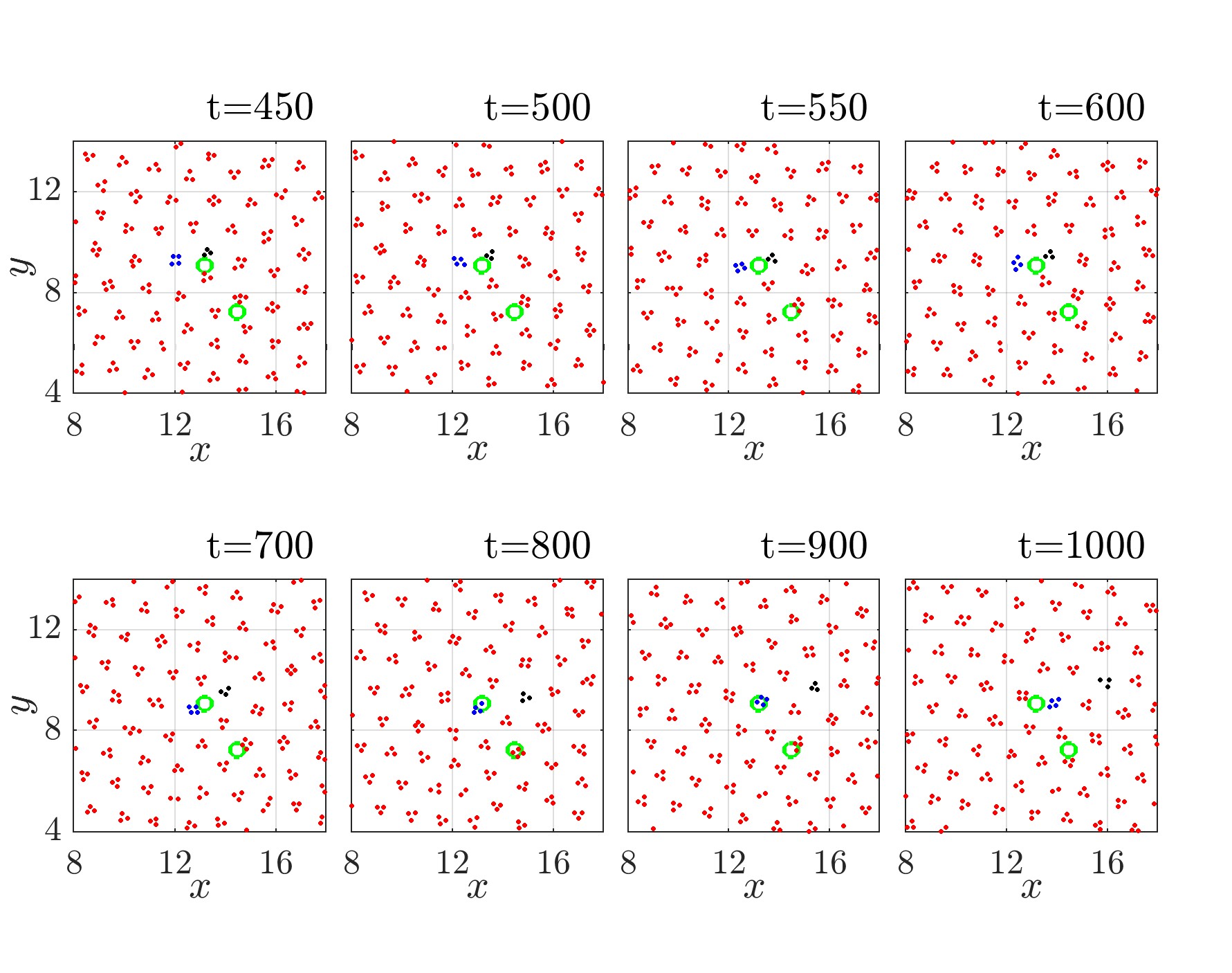}
\caption{
Example of a plastic neighbour-changing process for a dilute pinning case with $M=10$. The clusters can change their neighbours due to their polydispersity and consequently their mobilities. Here, the smaller black cluster depins faster and has a better mobility. As a result, when the blue cluster depins, the black cluster is no longer a neighbour of the blue cluster, inducing weak plastic flows.
}
\label{pw}
\end{center}
\end{figure}

Figure~\ref{pw} illustrates a neighbour-changing plastic flow process in the dilute pinning case. Here, the two relevant clusters are highlighted in blue and black. The blue cluster has four particles while the black cluster has three particles. The black cluster depins at around $t=450$ and afterwards the blue cluster arrives at the pinning center and is temporarily pinned. This pinning is stronger than that of the black one as there are more particles, and also importantly a smaller cluster has a higher mobility and is easier to diffuse around. As a result, although the blue cluster eventually depins also at around $t=1000$, the black cluster has already diffused away and is thus no longer a neighbour of the blue cluster. This process makes the flow weakly plastic. We emphasize, however, that such event does not occur frequently in dilute pinning; compare the time scale with Fig.~\ref{pf}.

To summarize this section,
all the depinning flows that we studied are plastic cluster flows.  Clusters tend to flow along pinning 
channels when pinning is dense. The effective polydispersity assists plastic flows in two aspects. First, it generates new forms of plastic flow processes such as clusters merging and breaking. In addition, their different mobilities can also induce neighbour-changing plastic flows when flowing through the pinning centers. These are quite distinct features from the depinning of ordinary non-cluster-forming glasses. 

\subsection{Dynamical reordering, critical exponent, and comparison with particle glass}

In this section we first demonstrate that cluster glasses upon depinning can undergo dynamical reordering when the driving force is increased. Next, we measure the critical exponent $\beta$ of the depinning transition. Finally, the cluster glass pinning strength is compared with that of ordinary particle glass, finding that cluster glass can have much stronger pinning effects than particle glass. This is potentially useful in superconductor applications.

\begin{figure}[htb]
\begin{center}
\includegraphics[width=\columnwidth]{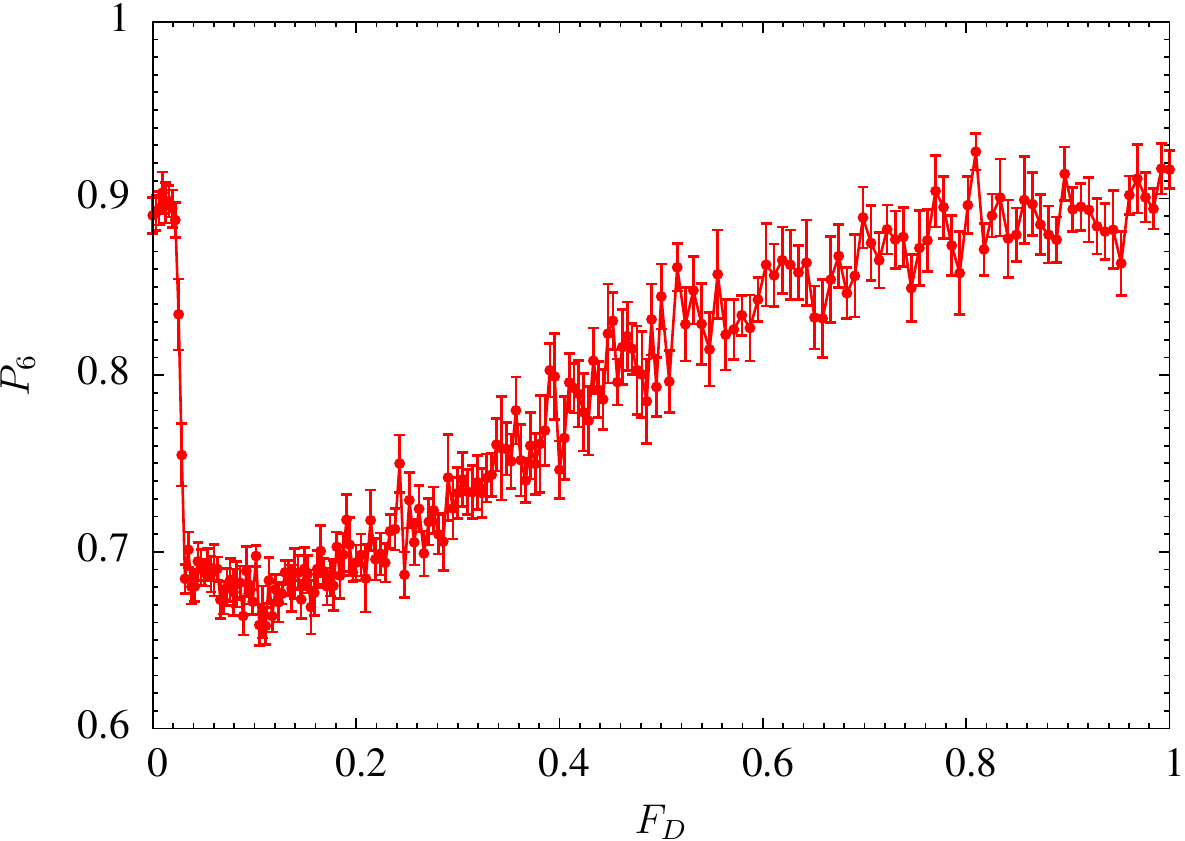}
\caption{
Cluster glass plastic flow exhibits dynamical reordering when the driving force is increased. Here the fraction $P_6$ of clusters with $6$ neighbouring clusters is shown as a function of the driving force for $M=160$. The fraction decreases significantly at the onset of the plastic flow around $F_D=0.02$, but it then gradually restores when $F_D$ is increased as the pinning effects are increasingly washed out. Note that this dynamical reordering appears to be a crossover rather than a genuine phase transition.
}
\label{P6}
\end{center}
\end{figure}

It is well known that the particle glass and the stripe pattern-forming systems can exhibit dynamical reordering as the driving force is increased \cite{RR:stripes,RR:depinning}. For example, stripes may break into pieces in presence of pinning disorder, but upon depinning stripes may reform when the driving force gets larger. Here, we investigate this phenomenon for the cluster glass. To study the dynamical reordering, we look at an order statistic $P_6$. This is the fraction of clusters having $6$ cluster neighbours in the asymptotic steady state as a function of the driving force. The result of $M=160$ is shown in Fig.~\ref{P6}. 
It is interesting that $P_6$ drops very sharply at the onset of the plastic flow, the critical force here is in good agreement with the depinning transition of Fig.~\ref{Drive}. This indicates that the system in the flowing phase is most chaotic when is just depins, and pinning and driving compete most strongly near the depinning transition. As the driving force increases, it is remarkable that $P_6$ is gradually restored and the effects of pinning are increasingly washed out; note that the force scale is larger than that of Fig.~\ref{Drive}. When $F_D \gtrsim 0.8$, $P_6$ is almost the same as that of the pinned phase. However, it seems that the reordering is by no means perfect, it does not achieve a perfect order for the largest force we studied, which is approximately $50$ times larger than the depinning force. In addition, the reordering occurs gradually and appears to be a crossover rather than a genuine phase transition.

\begin{figure}[htb]
\begin{center}
\includegraphics[width=\columnwidth]{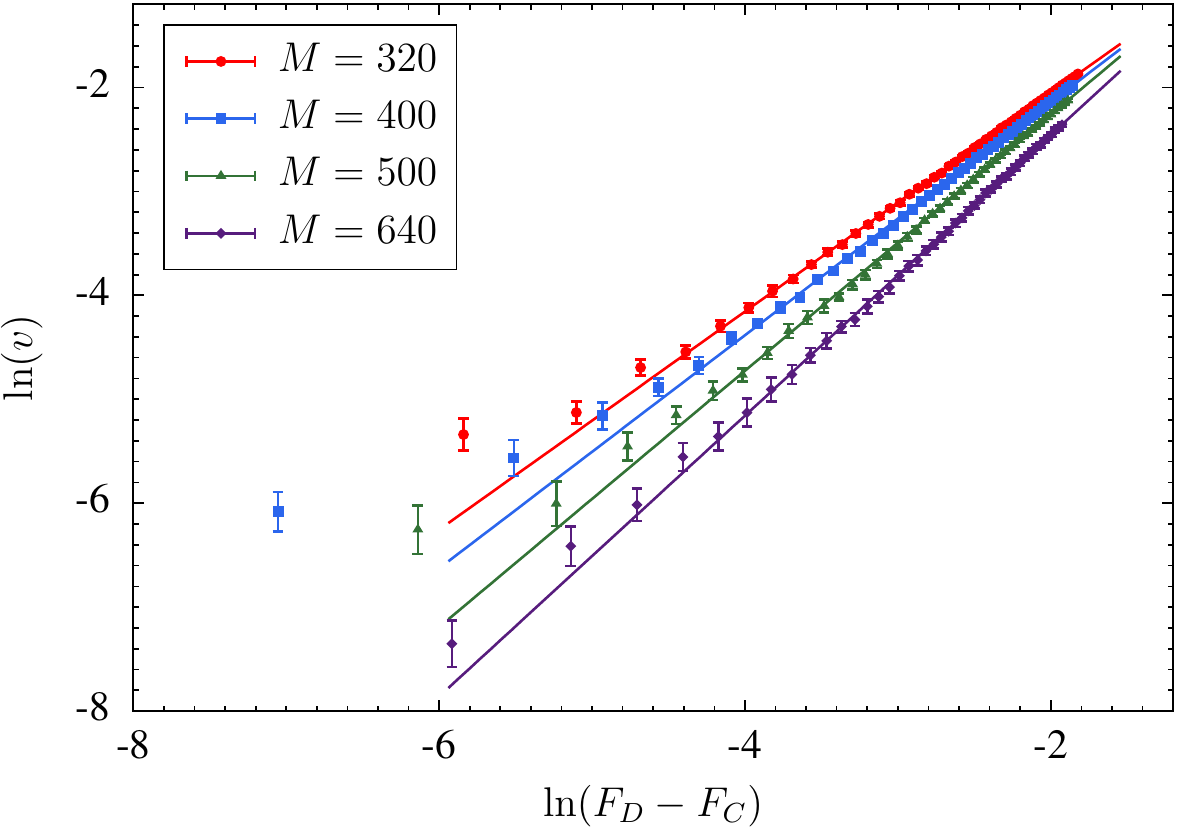}
\caption{
Power-law scaling of the velocity $v$ versus $F_D-F_C$ near the transition for $M=320, 400, 500$, and $640$. While the power-law behaviour appears to be reasonably good in all cases, the exponent or the slope becomes slightly larger with the increasing number of pinning centers, yielding $\beta=1.049(5), 1.121(5), 1.233(8)$ and $1.351(13)$, respectively. See the text for more discussions.
}
\label{beta}
\end{center}
\end{figure}

We now study the critical exponent $\beta$ of the depinning transitions. The critical exponent $\beta$ is defined as $v \sim (F-F_C)^\beta$ when $F \geq F_C$ in the vicinity of the critical force $F_C$. When the pinning is dilute, the range of $F_D$ where a power-law behaviour applies is limited; see Fig.~\ref{Drive}. By contrast, the depinning in the dense pinning regime has a much wider range of forces pertinent to scaling, and we therefore compute $\beta$ using the cases $M=320$ and $640$. We use a cubic function fit in the flowing phase but keep away from the transition to first locate the critical force $F_C$ and then make a power-law fit. In both cases, we find good power-law behaviours and the scaling plots are shown in Fig.~\ref{beta}. The two exponents are close but slightly different, $\beta=1.049(5)$ and $1.351(13)$, respectively. Adding two intermediate cases $M=400$ and $500$, we find that there appears to be a slow drift of the exponent $\beta$.

This exponent is unfortunately in general very poorly understood, and it is yet unclear whether there is a universality class for plastic depinnings. Our case appears to be even more complicated as our starting configurations are by design quenched glasses rather than slowly annealed states. Hence, it is not clear whether different pinning densities, corresponding to different degrees of glassiness, necessarily have the same critical exponent. More pinning centers introduce more defects into the quenched cluster glass, and therefore the depinning could be more plastic for a denser pinning, yielding a slightly larger exponent. 
On the other hand, our computed exponents also fall within the broad range of exponents 
found for plastic depinnings in many different systems \cite{RR:depinning}.
Therefore, this could be a finite-size effect or finite-temperature effect and there is a universality class, or alternatively quenched glasses may be different from annealed glasses. This intriguing question requires further systematic studies in its own right, and we shall not discuss it further here.

\begin{figure}[htb]
\begin{center}
\includegraphics[width=\columnwidth]{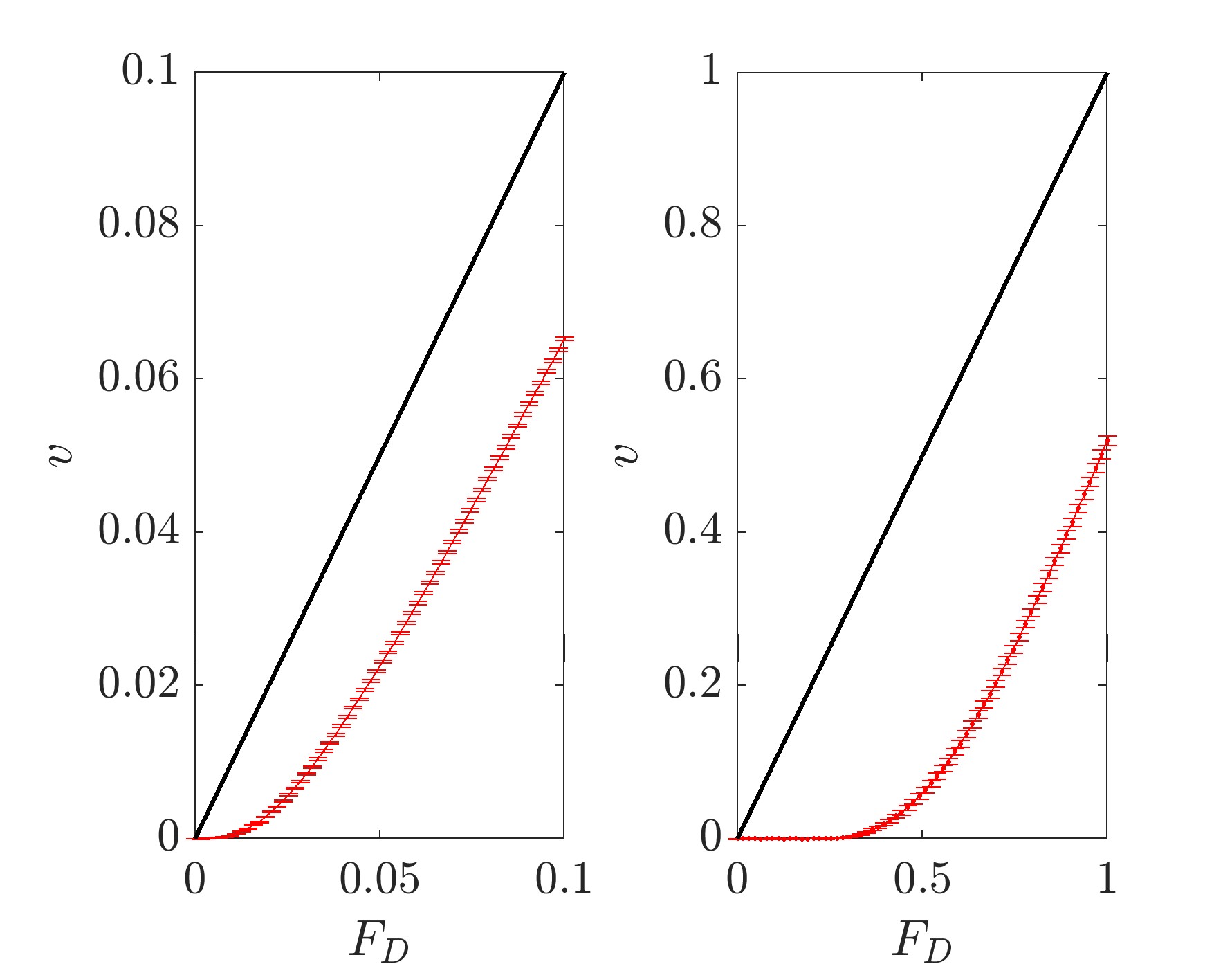}
\caption{
Velocity-force curve for particle glass (left) and cluster glass (right) for $M=320$ and $U_P=1$. The trap size is $\sigma=0.3799$, which is half the lattice constant of the particle crystal. Both glass formation and molecular dynamics are performed at their respective $T_C/2$. Note that the depinning force of the cluster glass is much larger than of the particle glass, suggesting that vortex clusters can have much better pinning properties than ordinary single vortices.
}
\label{ST}
\end{center}
\end{figure}

Finally, we 
compare the depinning forces of cluster glasses and particle glasses. To this end, we have simulated a group of particles at $N=1000$ and $n=2$ but interacting with a single Bessel function potential as shown in Fig.~\ref{UR}. The numerical setup is essentially the same for particle glasses, except that the temperature scales differ. Here, the transition temperature of the pure system is much smaller $T_C\approx0.01077$. Similarly, we prepare glasses and do molecular dynamics in presence of pinning at its own $T_C/2$.

We consider $M=320$ pins and 
$N=1000$ particles. We have similarly checked that the quenched states are particle glasses with $\phi_6 \lesssim 0.1$. 
There are two natural choices of the trap size $\sigma$ for a fair comparison, i.e., we can use either half the lattice constant of the cluster crystal or the particle crystal. For the same trap depth $U_p=1$, they yield wider or narrower traps, or weaker or stronger pinnings, respectively. Fortunately, they yield qualitatively the same conclusion. When wider traps are used, the particle glass has a very small depinning force that is not easily measurable, i.e., the traps are too wide to pin the particle glass effectively.
Hence we use smaller traps for \textit{both} systems, and the results are shown in Fig.~\ref{ST}.
It is found that the cluster glass again pins much better than the particle glass, the critical forces are approximately $F_C=0.01$ 
for particle glass and $0.3$ for cluster glass. There are many parameters including the interaction potentials that one can tune to explore this more systematically. 
However, considering that the depinning force scales are very different in the cases studied here, it is quite suggestive that cluster-forming type-1.5 superconductors can have much better pinning properties than system of ordinary vortices.

\section{Conclusions}
\label{cc}

We studied an ensemble of two-dimensional monodisperse particles interacting via a cluster-forming potential, relevant to vortices in type-1.5 superconductors in the presence of demagnetizing fields and layered type-1.5 systems. 
Vortices in such systems were previously demonstrated to form glasses in the absence of pinning, purely due to 
peculiarities of intervortex forces. 

We examined cluster glass formation following a thermal quench in presence of pinning centers of moderate depth, and the corresponding depinning transition. We find that weak and dilute pinnings have little effects on the interaction-dominated glass formation: the disordered
configuration adapts to the presence of dilute pinning sites, preserving the intervortex-interaction-driven effective polydispersity.

However, moderate, strong and dense pinnings can substantially assist glass formation, especially at intermediate temperatures, and indeed can entirely remove  order. For the pinning strength we have studied, the depinning transition is plastic and the cluster plastic flow exhibits features distinct from particle plastic flow. This includes cluster chaotic flow, clusters merging and breaking, and their different mobilities. Therefore, the effective polydispersity of clusters assists both glass formation and plastic flows. The cluster plastic flow is also found to exhibit dynamical reordering when the driving force is large.

Finally, a conclusion of practical importance is that  
cluster glasses can have significantly better pinning properties than systems
of non-cluster-forming particles. 
Since vortex pinning is important for transport properties of superconductors, it suggests that these
properties of  type-1.5 superconductors can be used in applications. Cluster-forming potentials can be engineered for example by layering systems \cite{meng16} or by using superconductors where there is an additional phase transition associated with the time-reversal symmetry breakdown \cite{carlstrom2011length,garaud2018properties}.

\acknowledgments 
We thank Cynthia Reichhardt for helpful exchanges on depinning transitions, and Alexander Zyuzin for helpful discussions.
W.W. and E.B. acknowledge support from the G\"{o}ran Gustafsson Foundation for Research in Natural Sciences and Medicine. E.B. acknowledges support from Swedish Research Council Grants No. 642-2013-7837, 2016-06122, 2018-03659, and the Olle Engkvists Stiftelse. W.W. also acknowledges support from the Fundamental Research Funds for the Central Universities, China. The computations were performed on resources provided by the Swedish National Infrastructure for Computing (SNIC) at the National Supercomputer Center in Link\"{o}ping, Sweden, and the High Performance Computing Center North (HPC2N) partially funded by the Swedish Research Council through grant agreement No. 2018-05973

\bibliography{Refs}

\end{document}